# Fast subsurface fingerprint imaging with full-field optical coherence tomography system equipped with a silicon camera

**Egidijus Auksorius**[*] **and A. Claude Boccara**
Institut Langevin, ESPCI ParisTech, PSL Research University, CNRS UMR 7587, 1 rue Jussieu, 75005 Paris, France

**Abstract**. Images recorded below the surface of a finger can have more details and be of higher quality than the conventional surface fingerprint images. This is particularly true when the quality of the surface fingerprints is compromised by, for example, moisture or surface damage. However, there is an unmet need for an inexpensive fingerprint sensor that is able to acquire high-quality images deep below the surface in short time. To this end, we report on a cost-effective full-field optical coherent tomography (FF-OCT) system comprised of a silicon camera and a powerful near-infrared LED light source. The system, for example, is able to record 1.7 cm × 1.7 cm *en face* images in 0.12 s with the spatial sampling rate of 2116 dots per inch and the sensitivity of 93 dB. We show that the system can be used to image internal fingerprints and sweat ducts with a good contrast. Finally, to demonstrate its biometric performance, we acquired subsurface fingerprint images from 240 individual fingers and estimated the equal-error-rate to be ~0.8 %. The developed instrument could also be used in other *en face* deep-tissue imaging applications because of its high sensitivity, such as *in vivo* skin imaging.

[*]E-mail: egidijus.auksorius@gmail.com

## 1 Introduction

### 1.1 Subsurface fingerprints

The quality of ~5% of all surface fingerprint images recorded with conventional fingerprint sensors is too low to be used for person recognition [1]. This can happen if a finger is dirty, wet, or dry. The surface fingerprint quality can also be compromised if a finger is affected by a skin disease, has superficial cuts, or other damage. Imaging such finger with a sensor that can only acquire surface images, like one based on frustrated total internal reflection (FTIR) principles will result in unusable fingerprint images. The above factors may have less of an effect to a finger structure that is below the surface of a finger due to the multiple layers of cells above the structure that can have a protective effect. In fact, the stratum corneum, is a thick layer on top of a finger composed mostly of dead cells that has a role of shielding live cell layers underneath it from mechanical and other damage [2]. The top skin layers composed of live cells are collectively called the viable epidermis and are mostly made of the stratum spinosum and thin layers of the stratum granulosum and the stratum basale. Some of these layers have essentially the same topography as the finger surface—the top of the stratum corneum—that is imaged to generate a conventional (surface) fingerprint image. However, it can be challenging to acquire high-contrast images of skin layers that are deep below the surface because of multiple scattering and absorption that light undergoes inside a finger. Optical coherence tomography (OCT) is a promising non-invasive technique for imaging such subsurface structure due to its ability to efficiently reject multiply scattered light through the coherence gating. OCT has been used in fingerprint sensing applications for imaging the so-called internal fingerprint since 2010 [3]. The term 'internal fingerprint' has been generally used to describe a fingerprint that is captured beneath the surface of the finger. In OCT-based fingerprint imaging studies it has been typically defined as a fingerprint that is captured at the papillary layer or dermis-epidermis junction. However, it is the viable epidermis, or more specifically, the stratum spinosum where most of the OCT signal is generated from beneath the skin surface [4-6]. The viable epidermis is interdigitated with the papillary layer and is at a similar depth, which could explain the confusion. In addition, histology of the 'thick skin' shows that the papillary layer should form primary and secondary ridges [2] but those are difficult to see in the cross-sectional OCT images. Therefore, it is more accurate to say that it is the viable epidermis that gives rise to the internal fingerprint image. Subsurface fingerprint is another general term used for a fingerprint acquired beneath the surface of a finger and we will use this term here in that sense. OCT can also be used to visualize sweat ducts inside the stratum corneum [7] and to image microcirculation of blood in the dermis [8] that might eventually be used for the liveness detection. Standard biometrics algorithms can be used on internal fingerprints since they also contain ridges and valleys, like conventional surface fingerprints. Specifically, minutiae extraction from points such as ridge endings and ridge bifurcations (splitting into two) can be performed to generate minutiae templates. The templates can be then used by the standard fingerprint matching algorithms for person verification/identification [1] by their internal fingerprints.

*1.2 Advantages of Full Field Optical Coherence Tomography*

Scanning OCT can acquire A-scans and B-scans (cross-sectional images) with very high speed; however, to generate a transverse (*en face*) image (C-scan), a volumetric (3D) data needs to be acquired. The *en face* image can be extracted from the 3D data but that could be computationally demanding, and moreover, the remaining volumetric data may be redundant. For a typical fingerprint image size of 3‑4 cm$^2$ and the spatial sampling rate of 500 dots per inch (dpi), an A-scan rate of at least 100 kHz is necessary to acquire an image in 1 second, which is a standard acquisition time in fingerprint imaging [1]. However, only one OCT study has been reported with the A-scan rate approaching the value of 100 kHz [9], whereas in the majority of studies the A-scan rate was only 16 kHz [3, 7, 10, 11], which resulted in a lengthy acquisition time and/or small fingerprint area. A short acquisition time helps to minimize image blur caused by involuntary finger movement, whereas a large imaging area ensures an adequate overlap between the fingerprints acquired at the enrolment and verification/identification steps. Furthermore, there is also a growing need for a larger sensor area for imaging multiple fingers [12]. Besides, higher sampling rates, such as 1000–2000 dpi or even higher, are necessary to image sweat pores and other "level 3" features that could be used to improve the fingerprint matching accuracy [13]. Recently developed Fourier domain mode-locked (FDML) lasers with the A-scan rates of up to a couple of MHz [14] allow very fast imaging. However, these are expensive and complex systems, mostly available in high-end research laboratories. For instance, commercial FDML laser source that can achieve a 1.5 MHz A-scan rate (Optores GmbH, Germany) costs over US$60,000. In contrast, full-field OCT (FF-OCT) [15] can use an inexpensive camera and a light source to acquire *en face* OCT images. FF-OCT is mainly used in applications requiring high isotropical resolution (<1 μm) and/or fast *en face* imaging [16-18]. The C-scan rate in FF-OCT is only limited by the frame rate of a camera, which is typically in the range of hundredths of Hertz or much more [19]. Therefore, a FF-OCT system with a megapixel camera can achieve the same equivalent A-scan rate as OCT system with a megahertz FDML laser source, but at a fraction of the cost. Of course, FF-OCT cannot compete with FD-OCT in the volumetric imaging rate since FF-OCT acquires only one *en face* image, whereas FD-OCT acquires a whole 3D dataset in a single scan. In addition, a pinhole in OCT systems helps rejecting multiply scattered light, and thus, use the available detection bandwidth more efficiently for the ballistic light detection.

*1.3  FF-OCT System with a Silicon Camera*

Previous attempts to use a silicon camera for imaging internal fingerprints with FF-OCT were not successful largely due to the low full well capacity (FWC) value, i.e., the number of photoelectrons that can be stored in a single pixel. The FWC value is important as it defines the system's sensitivity, i.e., the minimum detectable reflectivity $R_{min}$, which is proportional to $R/FWC$, with $R$ being the reflectivity of the reference arm [15]. Therefore, to achieve a sufficient imaging depth and thus image the internal fingerprints we have previously built an infrared FF-OCT system based on the InGaAs camera [20]. On the one hand it had a large FWC value (of 2 Me$^-$) and on the other hand it was sensitive in the infrared region (0.9 μm - 1.7 μm), which permitted deeper imaging because of the lower light scattering in that spectral range. However, the camera was slow (25 fps) and expensive (US$40,000).

In this paper, we demonstrate that internal/subsurface fingerprint images can be acquired with an FF-OCT system consisting of a novel silicon-based camera and an LED light. The silicon camera was designed to have a large FWC value (of 2 Me$^-$), like the InGaAs camera used previously, but with ×28 higher frame rate (720 fps versus 25 fps) and ×6.3 more pixels (1440 × 1440 versus 640 × 512). The greater speed and the number of pixels allowed increasing the FWC value further mathematically through pixel binning and/or frame averaging. The increase in the total number of detected photons per image by more than two orders of magnitude enabled capturing subsurface fingerprint images with higher contrast and field of view than in Ref. [20], despite using shorter light wavelength. In addition, the novel silicon-based camera enabled use of an inexpensive LED source that is more available, compact, efficient, and powerful compared with the halogen lamp used in Ref. [20]. The camera was also ×5 less expensive and ×4 more compact compared to the InGaAs camera. All the above enabled building a simple and cost-effective bench-top FF-OCT system that was used as a subsurface fingerprint sensor in this paper. The total cost of the system was below $15,000, including the camera, which is expected to decrease significantly with further developments and increased demand of the camera. Finally, to evaluate the biometric performance of the developed system a database of subsurface fingerprint images were acquired and analyzed.

## 2    Methods

*2.1 FF-OCT system*

The developed FF-OCT system, shown in Fig. 1, utilized LED (M850L3, Thorlabs) emitting 900 mW at 850 nm with a spectral bandwidth of 30 nm. Kohler illumination system was implemented for homogeneous sample illumination and stray light control by using singlet lenses

L1-L3 and diaphragms D1-D2. Light, collimated with a lens (L3), was sent through a beamsplitter that divided it equally into the reference and sample arms. Light coming back from the sample and the reference arm was recombined by the same beamsplitter and imaged onto the camera with the 1:1 magnification by combination of the objective (L4) and tube (L5) lenses. Both lenses contained a pair of achromatic doublets (f = 20 cm) arranged in the Plössl configuration to minimize the off-axis aberrations. The camera (Q-2A750-CXP, Adimec) was based on CMOS sensor (CSI2100, CMOSIS) that had over 2 million pixels (1440 × 1440), each 12 μm in size. The sensor was designed and its gain was set to achieve the maximum FWC value of 2 million electrons in each pixel.

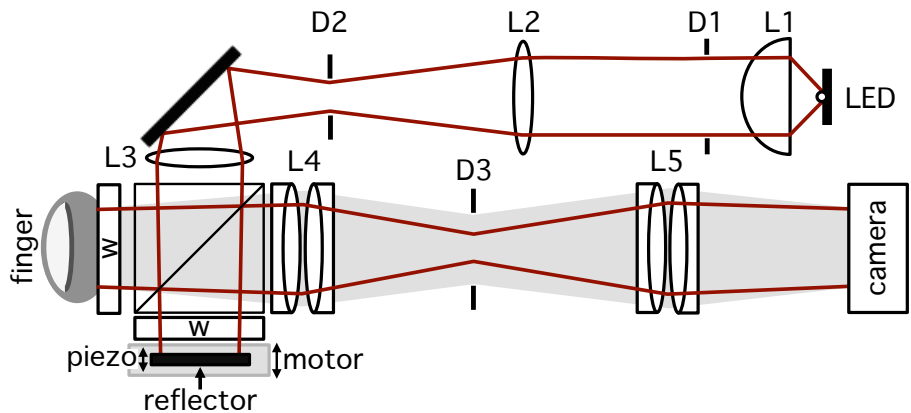

**Fig. 1.** Schematic of FF-OCT fingerprint system. L1 – aspheric singlet lens; L2 and L3: singlet lenses. L4: objective lens; L5: tube lens; D1: field diaphragm; D2: aperture diaphragm; D3: detection diaphragm; w: glass window. Red beam shows illumination and reference light, whereas gray beam represents scattered light coming from a sample (finger).

The camera could acquire images at the maximum speed of 720 frames per second (fps). Images from the camera were transferred with a CoaXPress frame grabber (Cyton-CXP4, BitFlow) that could sustain the maximum data transfer rate of 25 Gb/s. A sample area of 1.7 cm × 1.7 cm was imaged on the camera with the spatial sampling rate of 2116 dpi. Numerical aperture (NA) was set to 0.035 by limiting the detection diaphragm (D3 in Fig. 1) to 7 mm that resulted in the estimated transverse resolution of 16 μm, which was slightly undersampled by the camera. An antireflection-coated glass window, labeled (w) in Fig. 1, was used to press a finger against it in order to flatten it out and increase its stability during imaging. An identical window was inserted into the reference arm to match the spectral dispersion between the two arms. Light formed an interference pattern on the camera only when the optical path length difference between the sample and the reference arms was within the coherence length of the light source. The coherence length was measured to be of 10 μm in the air and of 7.5 μm in finger, estimated assuming that a typical refraction index of epidermis is 1.42. The reference arm was attenuated to 1% by either using a weak reflector or a mirror in conjunction with a neutral density (ND) filter. In the latter case a blank window of the same thickness and material was inserted in the sample arm for the dispersion compensation. This particular reflectivity of 1 % was chosen to match the total part of light being reflected and scattered from a finger when it is pressed on the window since the most optimal performance in terms of image quality is achieved when the power coming from both arms is equal. The reflector (a mirror or a weak reflector) was mounted on a piezo actuator that performed the phase shifting.

*2.2 FF-OCT Image Acquisition*

Following image acquisition and transfer to a computer, 20 images per phase were averaged before extraction of an FF-OCT image using a simple standard algebraic formula [15]. Since four phases were used to reconstruct the FF-OCT amplitude image the resulting total acquisition time was 0.12 s. That number of images was chosen as a compromise between the signal-to-noise ratio (SNR) and the artifacts occurring from the involuntary finger movements in FF-OCT images. Volumetric (3D) data were generated by stepping a motor in the reference arm and acquiring an *en face* FF-OCT image at each (axial) position. Sensitivity was measured by replacing a sample with a known reflectivity target – a mirror and ND filter that attenuated the signal to 1%. To achieve the maximum sensitivity, a sample (and the reference arm) was illuminated with 140 mW of light that put the camera's operation close to its saturation level – necessary to achieve the optimal performance. The sensitivity achieved by the FF-OCT system with the integration time of 0.12 s was estimated to be 93 dB from the following expression: $20 \times \log(I_{max}/\sigma) + 20$ dB, where $I_{max}$ is the signal amplitude recorded at the mirror surface, $\sigma$ – standard deviation of noise recorded far away from the mirror and 20 dB accounts for the 1 % attenuation in the sample arm. To increase the sensitivity further the spatial averaging in the transverse plane or, in case of 3D imaging, also in the axial direction was performed, as explained below.

## 3   Results and Discussions

### 3.1  Imaging Internal Fingerprints and Sweat Ducts

In this section we demonstrate that internal fingerprints and sweat ducts can be imaged with the developed FF-OCT system. To this end, a stack of FF-OCT images was acquired beneath the finger surface by stepping the reference reflector every 10 µm (that corresponded to ~7.5 µm in the finger) between the images. Fig. 2 (a) and Fig. 2 (b) show *en face* FF-OCT images derived by averaging 5 consecutive images over the depth range of (a) 70 µm – 100 µm and (b) 170 µm – 200 µm beneath the surface of the finger, which corresponds to the middle of the stratum corneum and the internal fingerprint, respectively. Averaging increased the total acquisition time to 0.8 s (that included the stepping) allowing the direct comparison (in section 3.2) with the images acquired with the FF-OCT system based on the InGaAs camera [20]. The image acquired from the inside the stratum corneum, displayed in Fig. 2 (a), clearly shows sweat ducts as distinct white formations.

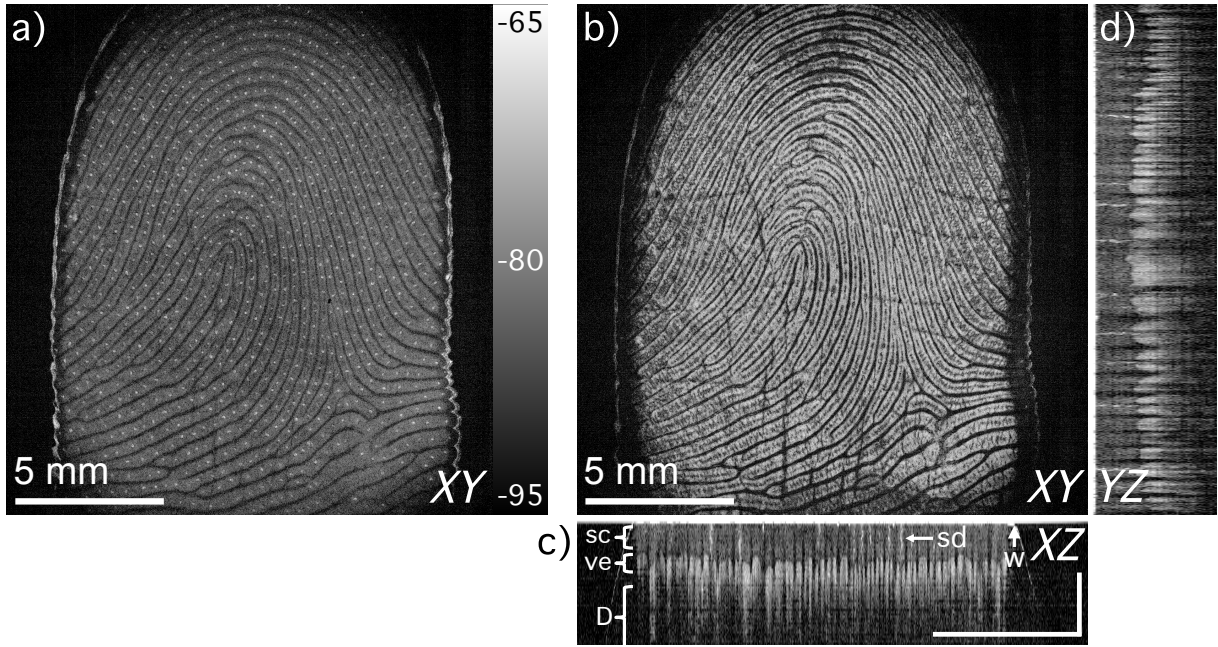

**Fig. 2.** Subsurface FF-OCT images. (a) Image derived from inside the stratum corneum (70 μm – 100 μm below the surface). It shows sweat ducts and low-contrast pattern of ridges and valleys. (b) Internal fingerprint recorded from the viable epidermis (170 μm – 200 μm below the surface). It shows high-contrast pattern of ridges and valleys. (c) Cross-sectional image *XZ* and (d) Cross-sectional image *YZ*. Mainly two distinct skin layers can be seen cross-sectionally: the stratum corneum (*sc*) and the viable epidermis (*ve*). Part of the dermis (*D*) can also be seen but it is not well contrasted. A sweat duct (*sd*) is pinpointed in the stratum corneum. Surface of the window (w), against which a finger was pressed, produces intense specular reflections seen as a thin bright line on top of the axial images. Five images were averaged along the following directions in the stack: *z* in (a) and (b), *y* in (c) and *x* in (d). All the images are displayed on the same logarithmic reflectivity scale in decibels – from *-95 dB* to *-65 dB*. Spatial sampling rate is 2116 dpi for the lateral (*XY*) images (no binning was performed). The scale bar for the cross-sectional images is 0.2 mm × 5 mm.

The appearance of sweat ducts might be different from sweat pores on surface because of contrast inversion and also because of surface artifacts having less of an affect on sweat ducts (compare Fig. 4 (a) and Fig. 4 (b), for example). It might, therefore, be easier to extract their features, such as location or shape, for use as a biometric characteristic [13]. Algorithms have been developed to identify individuals by their images of sweat pores [13, 21], and they will be adopted for images of sweat ducts in our future research. Most importantly, ridges and valleys can also be seen in Fig. 2 (a), which is a significant observation since it means that subsurface fingerprints can be acquired from inside the stratum corneum. Such a subsurface fingerprint and sweat ducts inside the stratum corneum can be seen in greater detail in Fig. 4 (b), which is a zoomed-in image displayed on a linear scale. An image of an internal fingerprint acquired from the viable epidermis is displayed in Fig. 2 (b), which shows the ridges and valleys with a clear contrast. The contrast is opposite to that of the conventional (surface) fingerprint, like one shown in Fig. 3 (a), because of the differences in the contrast mechanism (scattering versus FTIR). Fig. 2 (c) shows a cross-sectional image *XZ* of a finger that was derived by averaging over 5 successive cross-sectional images along *y* direction in the acquired stack. Cross-sectional image *YZ*, shown in Fig. 2 (d), is produced by such averaging along *x* direction. The averaging number was chosen to match that used in deriving the *en face* images in Fig. 2 (a)&(b). As shown in Fig. 2 (c), the main skin layers clearly distinguished with FF-OCT are the stratum corneum (*sc*), the viable epidermis (*ve*) and also part of the dermis (*D*).

*3.2 Comparison with the Previous Work*

The *en face* images in Fig. 2 can be compared with their counterpart images in Fig. 3 of Ref. [20], which were acquired with the infrared FF-OCT system based on the InGaAs camera. Note that the same integration time of 0.8 s was used for the both types of images, but in Ref. [20] images were displayed on a linear scale. Images reported here have a ×2.3 larger field of view and show higher contrast despite the use of ×1.5 shorter wavelength. Collecting ×180 more photons per second, as afforded by the silicon camera, counterbalanced the image degradation caused by the increased light scattering at the shorter wavelength. Furthermore, to achieve the same imaging performance in terms of the *en face* speed (~2 million pixels in 0.8 s) with a standard OCT system, a laser with the A-scan rate of ~2.5 MHz would be necessary, which is not currently available commercially.

*3.3 Fast Subsurface Fingerprint Imaging*

The FF-OCT system reported in this work allows fast internal structure *en face* imaging. For example, FF-OCT image of the internal fingerprint, shown in Fig. 3 (b), was recorded in just 0.12 s at the depth of 170 µm below the surface. Since the algorithms used for the minutiae extraction (see section 3.4) do not show accuracy improvement beyond the spatial sampling rate of 500 dpi we, therefore, reduced it from 2116 dpi to 500 dpi, which increased the sensitivity by 12.5 dB. Note, that the reduction in *dpi* is not strictly necessary and was not done on the images of the sweat ducts, shown in Fig. 2 (a), since that would remove its high-resolution details, as well as other level 3 features. Due to the sensitivity increase, the internal fingerprint in Fig. 3 (b) appears brighter than that in Fig. 2 (b) despite the shorter acquisition time.

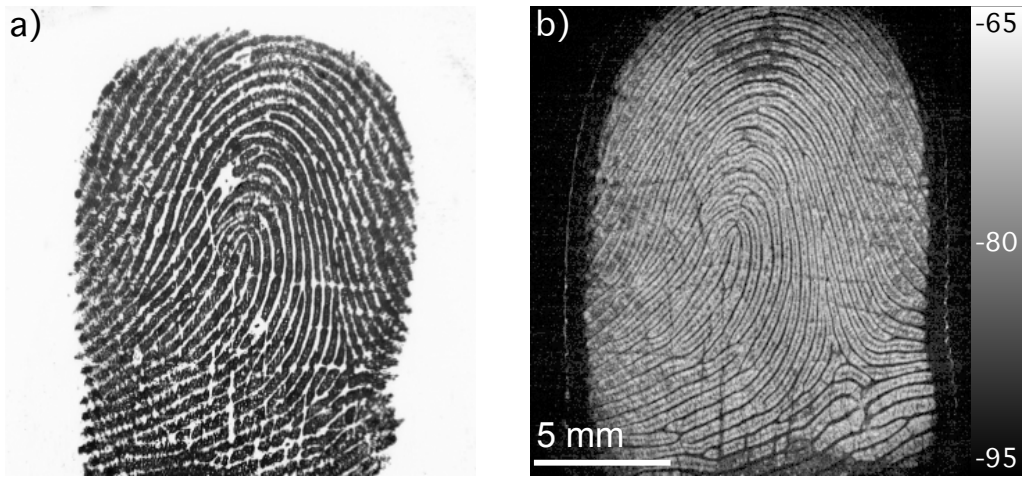

**Fig. 3.** (a) A conventional fingerprint image (displayed on a linear scale) acquired from surface (top of the stratum corneum) by the FTIR-based fingerprint sensor (MSO300, Morpho) with the spatial sampling rate of 500 dpi. (b) An internal fingerprint image (displayed on a *log* scale) acquired by the FF-OCT system in 0.12 s at the depth of 170 µm below the surface. To increase the sensitivity the spatial sampling rate was reduced from 2116 dpi to 500 dpi. The colorbar shows reflectivity in decibels.

The depth of internal fingerprints is known to change from finger to finger. For example, in this study we find that it varies approximately from 100 μm to 700 μm. Therefore, we cannot rely on a single imaging depth when a large number of fingers has to be imaged. In the simplest approach, *en face* images every 50‑100 μm could be acquired in the range of 100 μm – 700 μm, which, however, would extend the overall acquisition time to a couple of seconds. Also, a more careful image addition would be necessary for images acquired deeper due to the contrast inversion. For example, an image of an internal fingerprint recorded at the depth of its ridges, shown in Fig. 4 (c), will have an opposite contrast compared to the image recorded at the depth of its valleys, shown in Fig. 4 (d).

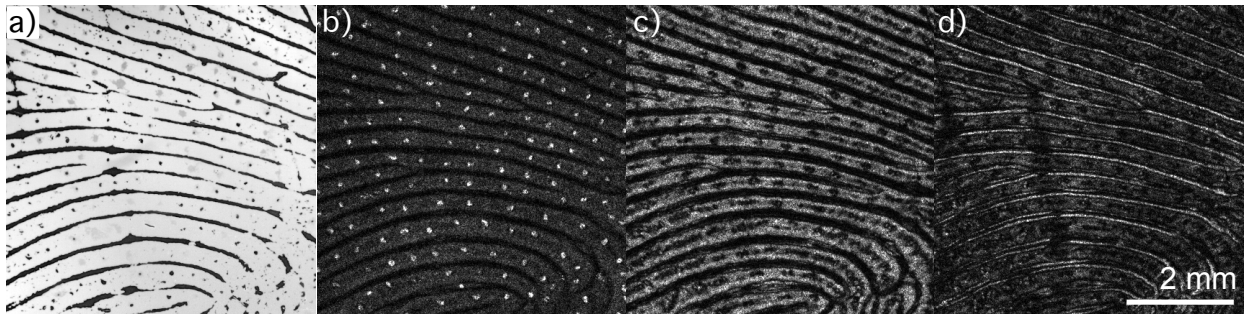

**Fig. 4.** Excerpts of FF-OCT fingerprint images acquired at the progressively increasing depth (from left to right): (a) top of the stratum corneum (surface fingerprint), (b) inside the stratum corneum, (c) ridges of the viable epidermis and (d) valleys of the viable epidermis. Sweat pores can be seen on the top of the stratum corneum (a), whereas sweat ducts – inside the stratum corneum (b). Note that the surface image (a) has an opposite contrast to that shown in Fig. 3 (a) due to the differences in the contrast mechanisms involved. Contrast inversion can also be observed between the images (c) and (d). Spatial sampling rate is 2116 dpi for all images (no binning was performed). Images were acquired at (a) 0 μm, (b) 77 μm, (c) 190 μm and (d) 280 μm with respect to the surface.

If two such images are added together it could reduce image contrast. An appropriate fusion algorithm can be developed adding the images together recorded at different depths with (or without) contrast inversion, similar to that reported in Ref. [22]. Alternatively, one can use *o*-FF-OCT method, suggested in Ref. [20], that relies only on two images to capture an internal fingerprint. However, since it also requires additional automation we did not pursue this method here. Instead, we rely on the observation made in section 3.1 that the ridges and valleys can also be seen inside the stratum corneum, like shown in Fig. 2 (a) and Fig. 4 (b). The features are practically identical to the ones seen on surface or in internal fingerprints but have somewhat lower contrast. Sweat ducts inverse their contrast in the internal fingerprint but that does not have a significant effect on image quality. The ridges and valleys in the stratum corneum and on the viable epidermis (internal fingerprint) do not disappear when an immersion liquid, such as water or oil, is applied on top of the finger. In other words, the fingerprint does not appear because of, for example, optical lensing effect produced by the surface topography. Therefore, acquisition of an internal fingerprint might be redundant if a fingerprint inside the stratum corneum can be captured. This can be useful, for example, when the internal fingerprint is located deep beneath the skin surface (say >500 µm) and therefore its quality could be seriously degraded due to the multiple light scattering and absorption. Since light is generally less scattered and absorbed in the stratum corneum due to its relative proximity to the surface, fingerprint images derived from there could be of higher quality despite its intrinsically lower contrast of the ridges and valleys. This utility of recording subsurface fingerprints from the inside of the stratum corneum needs to be further explored in the future by enrolling difficult fingers, such as ones from elderly people and manual workers. In this work we have empirically selected only two imaging depths: 140 µm and 210 µm that were chosen as a good compromise between the probability of capturing an internal fingerprint and the overall acquisition time. Also, the contrast inversion was not significant at these two particular depths. We find that at least a part of an internal fingerprint was captured in 70% of all fingers enrolled in the study here. The remaining 30 % of the images captured a fingerprint mostly from the inside of the stratum corneum. Therefore, we call this method a fast subsurface fingerprint imaging rather than internal fingerprint imaging since it generally contains information from the inside of the stratum corneum and the internal fingerprint. Although there is a certain degree of freedom in choosing the imaging depth, however, it should not be too close to the surface or too deep beneath it, because the subsurface fingerprints, on the one hand, could be affected by the surface artifacts when they are imaged too close to the surface but, on the other hand, they could be degraded by the multiple scattering and absorption when the imaging depth is deep. The acquisition procedure could be further simplified by recording only one subsurface image, which would, however, reduce the likelihood of capturing an internal fingerprint, and consequently, lower the fingerprint matching accuracy somewhat. It should not be surprising that fingerprint images can be acquired from the inside of the stratum corneum since other techniques also seem to use this information to enhance surface fingerprint images. Multispectral imaging (MSI) [23], for example, acquires a set of images at different wavelength, polarization and illumination configurations that probe the subsurface layers of a skin and generate a higher quality image after fusion with the surface fingerprint.

*3.4  Biometric study*

To demonstrate that such subsurface fingerprint images acquired with FF-OCT system can be used for the identification purposes we have recorded a database of 480 fingerprint images consisting of 24 individuals, 10 fingers per individual, and 2 images per finger. Each of those 480 images were produced by averaging the two subsurface fingerprint images captured at the depths of 140 μm and 210 μm in the total acquisition time of 0.3 s, as described above. Averaging the two FF-OCT images and reducing the spatial sampling rate to 500 dpi increased the sensitivity by 15.5 dB—from 93 dB to 108 dB. Fig. 5 (a) shows an example of such a subsurface fingerprint image that was acquired from the same finger as the one in Fig. 3 (b), except that the latter was acquired at the depth of 170 μm and displayed on the logarithmic scale.

The database contained volunteers from a wide age range and almost equal gender ratio. It had 4 individuals over 70 years of age but was dominated by ones from 20 to 40 years old. Informed consent was obtained from all the subjects allowing us to use their fingerprints for the research purposes. The volunteers were asked to put a finger in the center of the window and press it slightly. We found that varying the pressure did not change subsurface fingerprint images noticeably. Once all the fingers of an individual were imaged the acquisition session was repeated. For evaluation of biometric performance, the fingerprints recorded from the first session served as reference images and from the second – as test images. This resulted in 240 matching pairs or genuine attempts. To realize non-matching pairs (or imposter attempts), each of the 240 fingers were compared against the remaining 239 fingers, creating $240 \times 239 = 57,360$ non-matching pairs. Minutiae extraction and fingerprint matching were performed using standard biometrics software (Verifinger/MegaMatcher, Neurotechnology, Lithuania). Fig. 5 (a) shows an example of the extracted minutiae template (in red) overlaid with the FF-OCT image. Fig. 5 (b) shows the binarized image from which the minutiae extraction was actually performed. Some spurious minutiae could be identified in the image appearing mainly due to the ridge splitting in the internal fingerprint, which could be seen in greater detail in Fig. 4 (c). It is evident from the image that the splitting is caused by sweat ducts. Nevertheless, the spurious minutiae appear only in some parts of the image and are not seen in all the subsurface fingerprint images. We demonstrate below that fingerprint-matching results are adequate verifying that spurious minutiae problem is not significant. Nonetheless, we will remove the splitting artifacts through image processing in the future that would also make it more backwards compatible with the legacy (surface) fingerprints. Matching scores were calculated with the fingerprint matching software for all the matching and non-matching pairs. Selecting the matching score of 53 as a threshold resulted in 9 fingers being falsely rejected (3.75% of all the matching comparisons) but no fingers being falsely accepted (0% of all the non-matching comparisons).

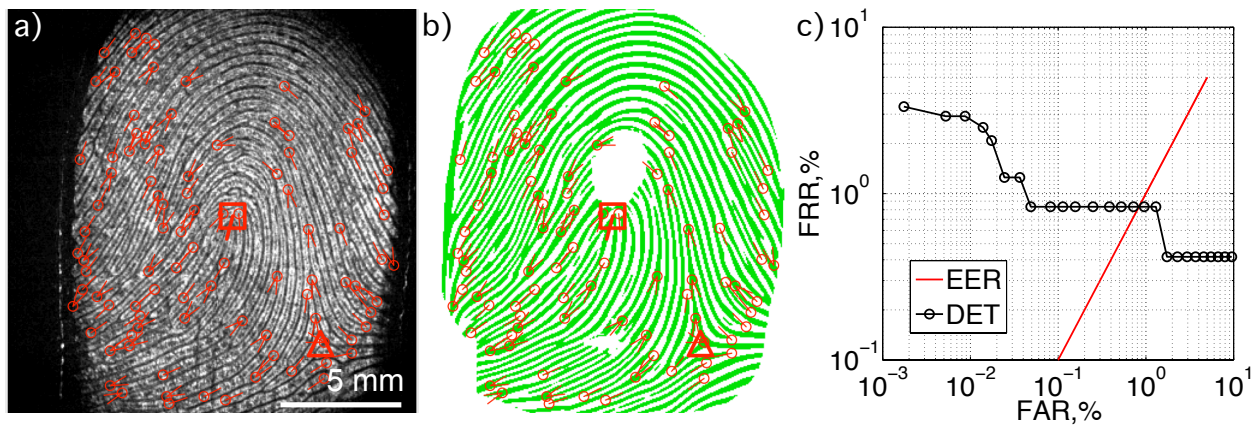

**Fig. 5.** (a) An example of a subsurface fingerprint image (acquired in 0.3 s and displayed with 500 dpi) that was one of the images used to evaluate the fingerprint matching performance. (b) Binarized version of the subsurface image shown in (a). Extracted minutiae points are shown in red in both images. (c) The DET curve on a *log-log* scale. EER curve (in red) is composed of equal FAR and FRR values.

In other words, the false rejection rate (FRR) was 3.75% at the false acceptance rate (FAR) of 0%. Fig. 5 (c) shows a detection-error tradeoff (DET) curve that plots the FRR and FAR values calculated for various other matching scores. The equal-error-rate (EER) that summarizes the DET curve in a single number was estimated to be ~0.8%, which is a point in Fig. 5 (c) where DET and EER curves cross. In this work we have captured fingerprint images from all the presented fingers and used them for fingerprint matching experiments. To enroll all the acquired fingerprint images with the Verifinger software the threshold value for the image quality had to be lowered from the standard 40 to at least 11. The image quality control was switched off because it is derived specifically for the conventional fingerprints and thus might not work properly for the subsurface fingerprints. Effectively, the failure to acquire (FTA) and the failure to enroll (FTE) rates were equal to 0%. Despite the enrollment of the images that were deemed to be of low quality by the software we have reported the EER value of ~0.8%, which compares well with those obtained by other fingerprint sensors. However, since the results obtained on different databases should generally be compared with care, it is therefore not straightforward to assess our results in the broader context. Also, there is lack of large subsurface fingerprint imaging studies with detailed performance analysis that provide at least the EER value. We could find only two such studies that are similar in size and depth of analysis – one recent OCT [11] and one MSI study [23]. The OCT-based study was performed with a non-modified pricy commercial OCT device and reported the EER value of 2.63% for the internal fingerprint matching. After fusing it with the surface fingerprints the EER value was improved to 1.25%. Although the acquisition time was not reported in the paper we estimated it to be at least 4 seconds from other reported parameters: an image size of 256 × 256 pixels and the A-scan rate of 16 kHz. That is at least an order of magnitude longer than the acquisition time of 0.3 s reported here at slightly better EER value. The MSI study reported similar to ours EER value of 0.8%. However, the EER values from the two studies should be compared with care since the fingerprint images were acquired at different depths and on different set of fingers.

## 4   Conclusions

In this paper we have reported on a cost-effective high-performing FF-OCT system and demonstrated that it can be used for fast subsurface fingerprint imaging. To this end, we first showed that the system is capable of imaging internal fingerprints and sweat ducts with high resolution, quality and speed, which constitutes a significant improvement to our previously reported system based on the expensive InGaAs camera. Furthermore, to demonstrate its biometric performance the system was used to image 240 individual fingers. Each finger was imaged in just 0.3 s, which was possible due to the specifications of FF-OCT sensor and the observation that the ridges and valleys can be recorded from the inside the stratum corneum making imaging of internal fingerprints not strictly necessary. Despite the short acquisition time, the equal-error-rate of ~0.8% was achieved. Since this prototype sensor did not have the ability to assess the fingerprint quality in real time, some of the recorded images could have been acquired with low quality due to, for example, incorrect interactions of the users with the sensor. Therefore, the EER value could have been even better had a proper image quality assessment been implemented at the fingerprint sensing stage, forcing reimage of the low quality fingerprints. This can be rectified in the future version of the sensor that would allow the user to record a fingerprint image again when the quality is low. In addition, the EER value could be further improved through image enhancement and a more elaborate fusion algorithm than just averaging of the two (or more) images, which would make the minutiae extraction more accurate. Although the database was relatively small and it lacked in difficult fingers, nevertheless, it was the biggest OCT-based fingerprint imaging study in terms of the number of fingers enrolled. This study should be viewed as a proof of principle that the developed FF-OCT system could be used to identify fingers by their subsurface fingerprints. The future research will entail the system's performance comparison against one of conventional sensors by recording fingerprints from the same set of volunteers with both types of devices. Furthermore, by including more difficult fingers we expect to highlight cases where FF-OCT sensor could outperform conventional sensors.

In addition to the reported results, the developed system can be of a more general interest. For example, higher equivalent A-scan rates in the wide field OCT settings currently could only be realized with expensive high-speed detectors, such as Phantom [19] or FastCam [17] cameras. However, the camera used in this work is at least ×10 less expensive but can achieve ~10 dB higher sensitivity due to its ability to read ~×10 more photons per second, albeit at lower equivalent A-scan rates. Thus, it might be preferred in applications such as deep tissue *en face* imaging, where the speed can be traded-off for higher sensitivity. For example, it could be used for imaging skin cancer and other dermatological diseases [24, 25].

The system can be further improved by using a light source with longer coherence length (>100 μm), such as a VCSEL array [26], which could increase the depth over which the signal is integrated, and thus, increase SNR, albeit at the expense of lower axial resolution. It could also facilitate the capture of internal/subsurface fingerprints in fewer images. Dark-field detection [27] could further improve the imaging depth by rejecting specular reflections originating from the window and other optical elements.

*Disclosure*

No conflicts of interest, financial or otherwise, are declared by the authors.

*Acknowledgments*

This work was funded by the European Union's Seventh Framework Programme (INGRESS project). The authors wish to thank Charles Brossollet for help with the image acquisition software development, Martynas Vilkas for discussions on Verifinger software and Kiran B. Raja for providing a code for the fingerprint matching experiments.